\documentstyle[11pt,paspconf]{article}

\begin{document}
\title{Tracing interactions in HCGs through the HI\altaffilmark{1}}
\author{L. Verdes-Montenegro}
\affil{Instituto de Astrof\'{\i}sica de
Andaluc\'{\i}a, CSIC,Apdo. Correos 3004, E-18080 Granada, Spain; lourdes@iaa.es}
\author{M. S. Yun}
\affil{National Radio Astronomy Observatory,\altaffilmark{2} P. O. Box
0, Socorro, NM 87801, USA; myun@aoc.nrao.edu} 
\author{B. A. Williams}
\affil{University of Delaware, Newark, Delaware, USA, baw@udel.edu}
\author{W. K. Huchtmeier}
\affil{Max-Planck-Institut f\"{u}r Radioastronomie, 
 Auf dem H\"{u}gel 69, D-53121 Bonn, Germany,
 huchtmeier@mpifr-bonn.mpg.de}
\author{A. Del Olmo and J. Perea}
\affil{ Instituto de Astrof\'{\i}sica de
Andaluc\'{\i}a, CSIC,Apdo. Correos 3004, E-18080 Granada, Spain;
chony@iaa.es, jaime@iaa.es}

\altaffiltext{1}{Based on 
observations made with the VLA operated by the National Radio Astronomy 
Observatory and on data taken using ALFOSC,
which is owned by the Instituto de Astrof\'{\i}sica de Andaluc\'{\i}a 
(IAA) and
operated at the Nordic Optical Telescope under agreement between IAA
and the NBIfA of the Astronomical Observatory of Copenhagen.}
\altaffiltext{2}{The National Radio
Astronomy Observatory is a facility of the National Science 
Foundation
operated under cooperative agreement by Associated Universities, 
Inc.}

\begin{abstract}
We present a global study of HI spectral line mapping for 16 Hickson Compact
Groups (HCGs) 
combining new and unpublished VLA data, plus the analysis of the 
HI content of individual galaxies.
Sixty percent of the groups show morphological and kinematical signs of
perturbations (from multiple tidal features to concentration of the HI in
a single enveloping cloud) and  
sixty five of the resolved galaxies are found to be HI deficient with
respect to a sample of isolated galaxies. In total, 77\% of the groups 
suffer  interactions  among all its
members which provides strong evidence of their reality.
We find that  dynamical evolution does not always  produce 
HI deficiency, but when this deficiency is observed, 
it appears to correlate with a high group velocity
dispersion and in some cases with 
the presence of a first-ranked elliptical.
The X-ray data available for our
sample are not sensitive enough for a comparison with the HI mass;
however this study does suggest a correlation between HI deficiency 
and hot gas since velocity dispersions are known from the
literature to correlate with X-ray luminosity.

\end{abstract}

\section{Introduction and data}

HCGs constitute a 
specially appropriate laboratory for the study of interactions among galaxies. 
Our aim is to understand how star formation, morphology and dynamics are 
affected by the environmental conditions, and our approach is 
multiwavelength (e.g. Yun et al 1997, Verdes-Montenegro et al 1998). 
Since the atomic hydrogen is a sensitive tracer 
of tidal interaction (e.g. Hibbard \& Van Gorkom 1996, Hibbard 1999), 
we have performed VLA mapping 
of a large number of HCGs which added to the previously published 
ones (Williams \& Van Gorkom 1988 - WVG88, Williams, Mac Mahon and Van
Gorkom 1991 - WMVG 91, 
Williams \& Van Gorkom 1995 - WVG95, Williams et al 1997 - W97) and 
lead to a total sample of 16 groups. In two cases (HCG\, 31 and HCG\, 
79) we have obtained higher spatial and spectral resolution data 
which have helped to improve our understanding of these systems.
All the HI data herein are analyzed together for the 1st 
time and provide a sample to test previously proposed models
of compact groups as well as a detailed database for hydrodynamical 
simulations.
HCG\, 2, 18 and 54 are considered apart since  the first one is
 a triplet, and the others likely have less than 4 galaxies.

\section{HI content and distribution}

\begin{figure}
\plotfiddle{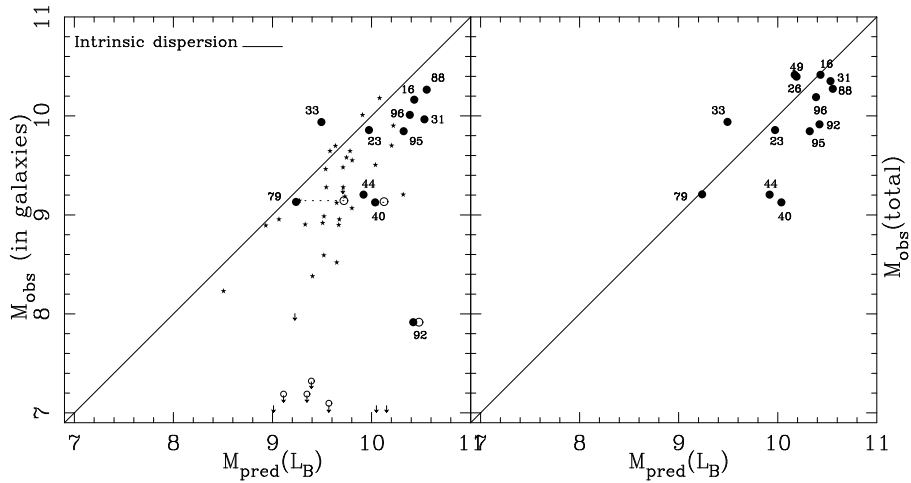}{5.5truecm}{0}{75}{75}{-190}{-220}
\caption{a) Observed against predicted HI mass in the disks of the 
individual galaxies
(stars for detections, arrows for upper limits, and empty
circles for S0s). The filled circles represent the 
HI mass of HCGs considering only the gas in the disks of the galaxies
and  excluding S0 types. The connected open circles show the result when they 
are included. The the difference is only significant for HCG\, 79.
b) Total HI mass detected in each group plotted against predicted HI mass.}
\label{fig-1}
\end{figure}

We have measured the HI mass associated with the main body of the galaxies
as well as that found in gaseous tidal features, except for HCG\, 26
(WVG88) and HCG\, 49, where the emission is found in a large envelope
containing all group members and therefore cannot be separated. The HI
content of the spiral and lenticular galaxies has been 
compared with the predicted one for their optical luminosity 
(Fig. ~\ref{fig-1}a) 
 based on the  relationships obtained by Haynes \& Giovanelli (1984) for
a sample of isolated galaxies. 
We note that HCG\, 33 has only a spiral
member, which has a bright star on the top, hence an accurate determination of 
L$_B$ cannot be obtained. 
Disks with a normal HI content
are found  but most galaxies have
significant amounts of missing gas, and this is evident in spite of 
 the intrinsic dispersion in the HI content ($\sigma$ $\sim$ 0.25 in log). 
None of the lenticular galaxies (empty circles) in this sample have
been detected, but 
since they usually show a wide range of HI
content, we have checked that their inclusion does not affect
significantly our conclusions.
The total mass expected for a group is also plotted, and has been 
calculated as the sum of the expected values for each member.
The use of the summed L$_B$ to predict a total HI mass would produce 
an artificial 
systematic deficiency in the log of 0.2 due to the non-linearity of the
M(HI) - L$_B$ relation.
We have plotted in Fig. 1b the total detected HI mass, i.e., including
gas external to the galaxies. Comparison of both panels tell us whether
 the missing HI is located in tidal features or just disappeared. This 
together 
with the morphology and kinematics of the HI show
the existence of  different distributions that we describe next.

\subsection{Most of the HI mass within the galaxies}

HCG\, 23, HCG\, 33 and HCG\, 88 have more
than 90\% of the HI mass associated to the disks and show little or no
signs of interaction (WVG95, W97). 
HCG\, 88 constitutes however a very interesting case as it might be 
considered a filament seen in
projection along the line of view. 
The low 
velocity dispersion of this quartet ($\sim$ 30 km/s) together with its 
high degree of isolation (De Carvalho et al 1997) contradict the chain
alignment hypothesis. Consequently we think that 
HCG\, 88 is a good example of a physically dense group in a 
very early stage of interaction.

\subsection{Significant HI mass in tidal features}

\begin{figure}
\plotfiddle{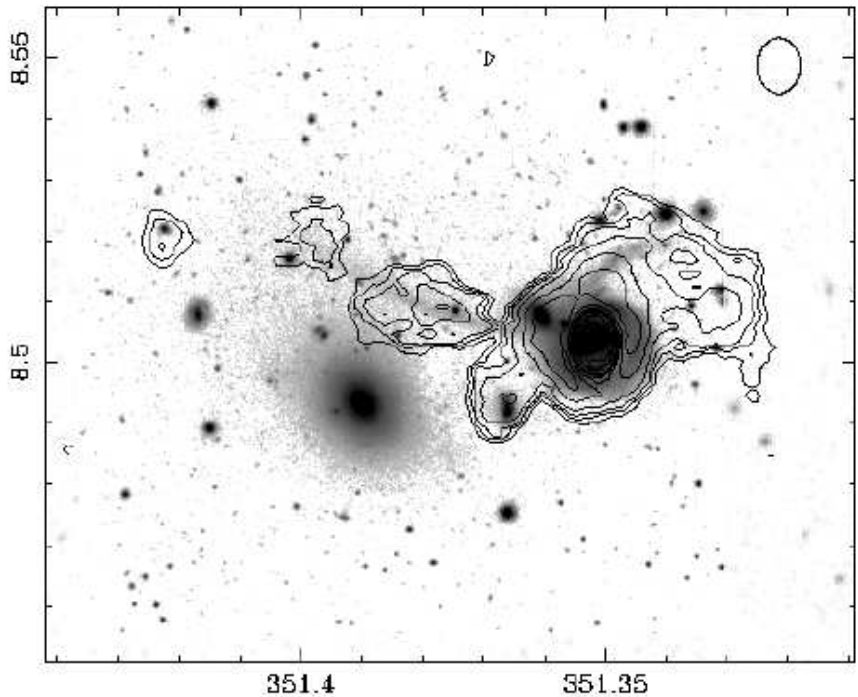}{7cm}{0}{70}{70}{-205}{-140}
\caption{R image of HCG\, 96 obtained at the NOT (Nordic Optical
 Telescope).
 The contours correspond to the HI column density 
distribution.}\label{fig-2}
\end{figure}

HCG\, 16, 31 and 96 show most of the missing gas 
in numerous tidal features, indicating that multiple interactions are 
taking place. 
In HCG\, 96, 30\% of the  HI 
mass is  located in two intense tidal tails 
plus a bridge. 
H96b is a bright elliptical with an optically detached core 
(Verdes-Montenegro et al 1997) plus an extended envelope brighter in 
the direction of the tidal tails (Fig.~\ref{fig-2}). 
Tidal features were also reported for HCG\, 16
from D array data by W97 and our new C array data resolve the emission  
indicating that a 40\% of HI is distributed in two intense bridges and 
two tails.
Consequently there is no room for 
diffuse X-ray emission in this group. A result that is inconsistent 
 with the conclusion reached by Mamon et al
(this conference).
The most extreme case is that of HCG\, 31 where 60\% of the  gas 
is located in 4 tidal tails and 1 bridge
(WMVG91, Del Olmo et al 1999). 
HCG\, 79, the densest group in the HCG catalog, could be 
included in this category since its only spiral member shows  
a tidal tail smoothly connected in velocity with the 
galactic disk. Considering the two anemic lenticulars,
 the group  is strongly HI deficient
 and  might be better placed in the next category, little HI coupled with
the galaxies. 
The optical diffuse envelope that contains the whole
group (Sulentic \& Lorre 1983) suggests that this configuration is 
the product of a much older
perturbation; yet, the HI morphology indicates a subsequent and more
recent interaction, i.e., tidal tail, that has not yet perturbed the HI
disk of the sole spiral member of the group. The presence of the diffuse
optical envelope makes it more plausible that the anemic lenticular
members were once spiral galaxies stripped of their gas during the
these earlier interactions. 

\subsection{Little HI coupled with the galaxies}

\begin{figure}
\plotfiddle{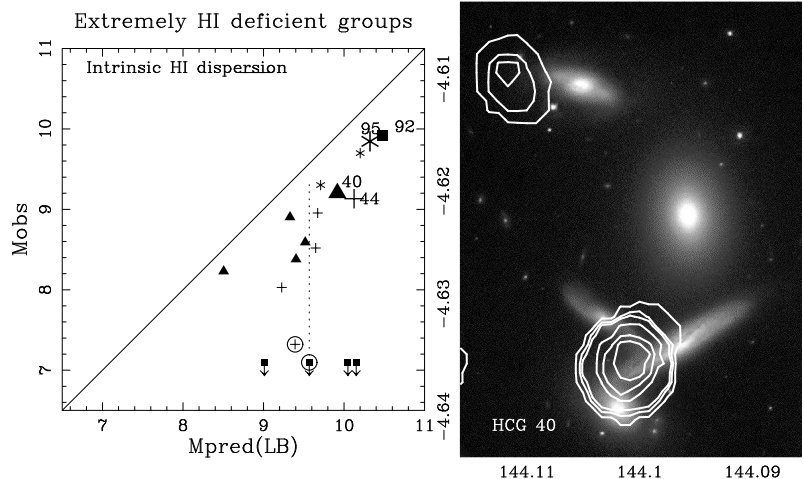}{5.7cm}{0}{70}{70}{-195}{-205}
\caption{a) Observed against predicted HI masses for the most HI deficient groups. 
Big symbols correspond to group masses and small
ones to individual masses. The open circles correspond to
S0 galaxies and arrows to upper limits. The dotted lines mark the 
location of each  galaxy in HCG\, 92 if all mass detected, located in external
features, would be assigned to it.
b) J \& K' band image of HCG\, 40 taken with Subaru where
the contours corresponding to the HI column density distribution are 
plotted.}\label{fig-3}
\end{figure}
\subsubsection{HI deficient groups.} HCG\, 40, HCG\, 44 (WMVG91), HCG\, 
92 (W99) and 
HCG\,
95 (Huchtmeier et al 1999) have an HI deficiency of 70 to 90\%, 
as found by Williams and Rood (1987) and
Huchtmeier (1997) from single dish observations. Our maps show that this
deficiency is due to all galaxies in 
the group (Fig. 3a). In the case of HCG\, 92 the 8 $\times$
10$^9$M$_{\odot}$ of HI detected are fully located in several clouds and
tidal features (W99), so we were not able to
discriminate from which members the gas was missing. However the multiplicity
of features plus the small amount of detected HI strongly suggest
that the missing gas was related to most if not all galaxies in the group.
In Fig. 3b we show the most striking mapped case, HCG\, 40, where
only one half of the disk of H40c is detected in HI plus a small 
cloud to the eastern side of H40d.

\subsubsection{Single HI cloud.} The HI towards HCG\, 26 (WVG88) and HCG\, 49
constitutes a single envelope with a velocity gradient
decoupled from the individual galaxies. In the case of HCG\, 49
the cloud is round-shaped, with a velocity gradient of $\sim$ 250~km~s$^{-1}$ 
containing 4 well differentiated galaxies with a velocity dispersion of 
34~km~s$^{-1}$  and a total optical diameter of 35 kpc. This cloud constitutes a 
challenge for dynamical models due to the coexistence of separated optical
morphologies and a global common kinematics.

\section{Discussion}

HCG galaxies respond to their environment at different levels: 
60\% have redistributed the atomic gas and  65\% have a lower
than expected content given the intrinsic dispersion of this quantity.
Since the gas is missing from all galaxies in the group it implies
that all of them are interacting with each other and/or with an intragroup 
medium. 
The total number of groups perturbed in one of these two ways amounts to
77\%.
 Both kinds of perturbations do not 
coexist in all cases: we find HI deficient groups for which most 
HI is found in tails and external clouds (e.g. HCG\, 92) while the groups 
embedded  in a single HI cloud have a normal HI content. 
The evolution from mild interactions to the generation of
multiple tidal features can be well understood as an extrapolation of  
interacting pairs. Formation of a single cloud 
with a coherent kinematics implies a larger degree of evolution, 
and it might be related to the fact that
these groups are mostly composed of dwarf galaxies (around 15 kpc in diameter)
which tend to have more extended HI disks, as suggested 
by Bosma (priv. comm.).
HCG\, 31 is a promising candidate to form a large
envelope due to its present HI distribution and kinematics together with
the small size of its galaxies. 

From our data, we point out two possible causes for the observed 
HI deficiency. One is the presence 
of a first ranked elliptical which could be the case for  HCG\, 40 and 95,
since they constitute a deep potential well 
that can accrete gas (V\'{\i}lchez \& Iglesias-P\'{a}ramo 1998). 
The second one can be the presence of hot gas, as in HCG\,
92 where tidal features contain all the detected gas which 
anticorrelates with a ridge of X-ray emission (Pietsch et al 1997).
The available X-ray data  for our HI sample
are mostly upper limits  therefore analyzes of the X-ray
correlations are limited.
We have found a correlation between HI deficiency 
and the groups´  velocity dispersion. 
The velocity dispersion of a different but larger sample of HCGs
correlates with X-ray emission (Ponman et al 1995).
This is in the sense that the deficiency increases with the
X-ray emission.
Gas accretion from giant ellipticals, or a hot medium 
may compete in the production of HI deficient groups.
These factors will be studied further in a subsequent paper.

\acknowledgments
LVM acknowledges interesting discussions with J. Sulentic,
R. Sancisi, E. Athanassoula and A. Bosma. LV--M, AO and JP are 
partially supported by DGICYT (Spain) 
Grant PB96-0921 and Junta de Andaluc\'{\i}a (Spain).

\end{document}